**Проучване и анализ на мнението на работодателите,
относно необходимите умения, които трябва да притежават студентите
в областта на уеб програмирането**
Йордан Калмуков

**Research and Analysis of Employers' Opinion on the Necessary Skills
that Students in the Field of Web Programming Should Possess**
Yordan Kalmukov

**Abstract:**
In the era of artificial intelligence (AI) and chatbots, based on large language models that can generate programming code in any language, write texts and summarize information, it is obvious that the requirements of employers for graduating students have already changed. The modern IT world offers significant automation of programming through software frameworks and a huge set of third-party libraries and application programming interfaces (APIs). All these tools provide most of the necessary functionality out of the box (already implemented), and quite naturally the question arises as to what is more useful for students – to teach how to use these ready-made tools or the basic principles of working and development of web applications from scratch.

This paper analyzes the results of a survey conducted among IT employers, aimed to identify what, in their opinion, are the necessary technical skills that graduating students in the field of Web Programming should possess in order to join the company's work as quickly and effectively as possible.

**Keywords:** web programming; frameworks; built-in functions and APIs; students' skills; employer requirements; usage of AI (ChatGPT, Claude, DeepSeek) in education and work.

**For contacts:** Assoc. Prof. Yordan Kalmukov, PhD, University of Ruse, jkalmukov@uni-ruse.bg

## INTRODUCTION

Web programming or web development is a popular fast-growing IT area that is considered to be part of the basic/mandatory programming skills all software engineers should have. The current trend in software development is that most applications became responsive web applications which can work on any desktop, portable, mobile or even embedded device. The biggest advantage of web applications is that they are available anytime (24/7), anywhere in the world, on any device that has browser installed. According to Real-time statistics on Developers [1], about 42% of all software developers are web developers. That is why all university software-related courses/specialties contain at least one, but usually more, subjects in Web programming.

In the era of artificial intelligence (AI) that can generate programming code in any programming language, write texts and summarize information, it is quite obvious that the requirements of employers for graduating students have already changed a lot [2, 3, 4]. Moreover, the modern IT world offers significant automation of programming through software frameworks and a huge set of third-party libraries and application programming interfaces (APIs). They provide most of the necessary functionality out of the box (already implemented), and quite naturally the question arises as to what is more useful for students – ***to teach how to use these ready-made tools*** or ***the basic principles of working and development of web applications from scratch***.

*Conducting a survey among employers and students*

To answer the abovementioned important question objectively, I have developed and conducted a survey among employers and final year students. It consists of 14 questions grouped in the following topics:
- ✓ Importance of algorithmic and logical thinking for graduates;
- ✓ Working with built-in functions and APIs vs. implementing basic algorithms from scratch;
- ✓ Teaching frameworks vs. basic principles and writing code from scratch;
- ✓ Using artificial intelligence (AI) tools during teaching/training, during work in a real environment, and during the exam – when it is acceptable and when it is not.

Respondents:
- ✓ 19 representatives of employers/companies in positions: "Manager" and "Team Lead". The companies include international corporations, large Bulgarian companies, and large and small local companies.
- ✓ 10 graduating students (bachelor and master's degree).

*Results and discussion*

Even the first question (fig.1.) reveals significant difference in the employers' opinion in respect to the students' one. Asking how important the algorithmic and logical thinking is for graduates, 89% of employers state that this is the most important thing that distinct good from mediocre programmer. However, almost half of the students do not think this is the most important thing they should learn at the university.

Q1. In your opinion, how important is it for graduates to have algorithmic and logical thinking?

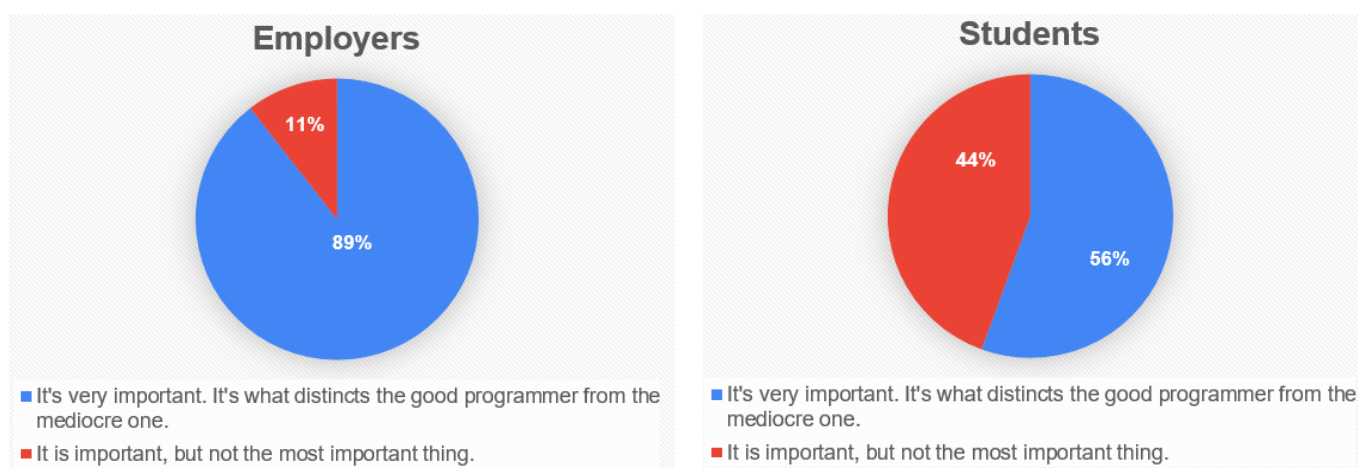

Fig. 1. Question 1 and percentage distribution of answers of employers and students

The next couple of questions (table 1) survey whether students should implement some basic algorithms for which all programming languages offer built-in functions and functionality that could be used as a black-box from external APIs. Majority of respondents state students should be able to implement basic algorithms, although they can use them as a ready-made built-in functions.

Table 1. Questions about the usefulness of implementing basic algorithms rather than using built-in functions or APIs

| Q2. In your opinion, should graduates know and be able to implement basic algorithms for which there are built-in functions in most programming languages?<br>(e.g.: finding the minimum and maximum element, arithmetic mean, sorting, searching in space, etc.) | | |
|---|---|---|
| | **Employers** | **Students** |
| Yes, mandatory | 48 % | 22 % |
| It is desirable | 26 % | 78% |
| Rather not, since there are built-in functions | 26 % | |
| | | |
| Q3. Should students learn (and practice) how to implement certain functionalities if external APIs exist that provide them out of the box? (e.g.: full-text search, basic data processing and analysis, etc.) | | |
| | **Employers** | **Students** |
| **Yes, it's better** to be able to implement them and adapt them to their specific task | 89 % | 67 % |
| **Yes, but only** if the services provided by external APIs are expensive | | 33 % |
| **No, there's no point** in learning such things, since they can be used from external APIs. | 11 % | |

Asking if it is important that graduates know at least one framework for the server-side programming language, they study, both students and employers state it is mandatory or important, however they also indicate that graduates should be able to develop web applications from scratch as well, without using frameworks (table 2).

Table 2. Frameworks vs writing code from scratch? Seems both are important.

| Q4. How important is it for graduates to know in details at least one framework for the given server-side programming language they are studying? | | |
|---|---|---|
| | **Employers** | **Students** |
| **It is mandatory** | 37 % | 44 % |
| **It is important**, as long as there is time and it does not interfere with mastering the relevant programming language | 42 % | 56 % |
| **It's not that important**. It's more important that they know the programming language itself well. | 21 % | |
| | | |
| Q5. In your opinion, should graduates be able to create complex web applications in a given programming language without the use of frameworks or other tools? | | |
| | **Employers** | **Students** |
| **Yes, of course**. Frameworks are useful, but programmers should be able to work without them | 47 % | 33 % |
| **It is desirable**, but if they can work with a framework, it is not necessary to know the programming language in detail | 37 % | 56 % |
| If they can create complex applications with a framework, **it does not matter** if they know the programming language at all | 16 % | 11 % |

Although both, writing code from scratch and teaching frameworks, are important it is not easy to cover them both, since the tutorials of a single subject are 24-26 hours (2 hours per week). So the question arises, due to these limitations, which is better – to teach how to use frameworks or basic principles of web applications and writing code from scratch? Fortunately and quite reasonably both students and employers state that it is better to have a good knowledge in the programming language itself, rather than in its framework (fig.2).

Q6. Yet, if you have to choose (due to the limited number of hours), what is more important: knowing the programming language in details or one of its frameworks?

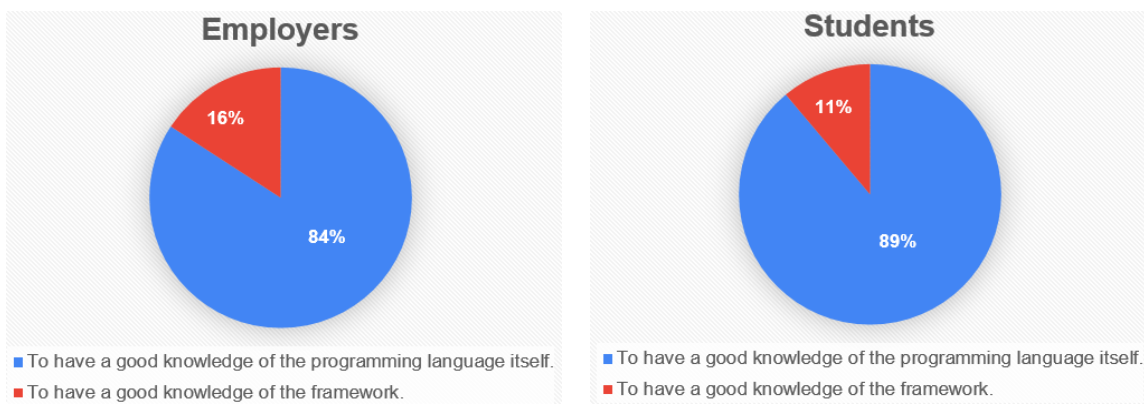

Fig. 2. What is more important: knowing the programming language in details or one of its frameworks?

The survey continues with questions related to client-side frameworks; SQL queries vs. ORM frameworks; types of exams suitable for web programming; types of coursework and so on. Due to the page limitation on this paper, the analyses of these questions will be omitted. Just at a glance – both employers and students require at least basic knowledge of client-side frameworks; to know both SQL and ORM frameworks; and think that practical exam on a computer is the most suitable for web programming.

Now let's focus on another group of questions – the ones related to the use of Artificial Intelligence (AI) in education. Asked, if AI tools should be actively used in education, the majority of employers and students response that they would be very useful (table 3).

Table 3. Knowing how to use tools and IDEs with integrated AI is found useful by both employers and students.

| Q11. Should the learning process include and rely on tools and development environments with integrated AI (artificial intelligence)? | Employers | Students |
|---|---|---|
| **Absolutely**. AI is leading to significant automation of routine programming tasks. Students need to know how to use it. | 21 % | |
| **It would be useful**. Even if AI is not actively used in the educational process, students should know how it could make their live as developers easier. | 58 % | 89 % |
| **Rather not**. Let students learn to program on their own, develop the necessary algorithmic thinking. | 21 % | 11 % |

Probably the most important question from the AI-related group is whether it is acceptable that graduate students can write code with the help of artificial intelligence, but cannot do it alone (fig.3).

Q12. How acceptable is it for a graduate student to be able to write code with AI (artificial intelligence), but not alone?

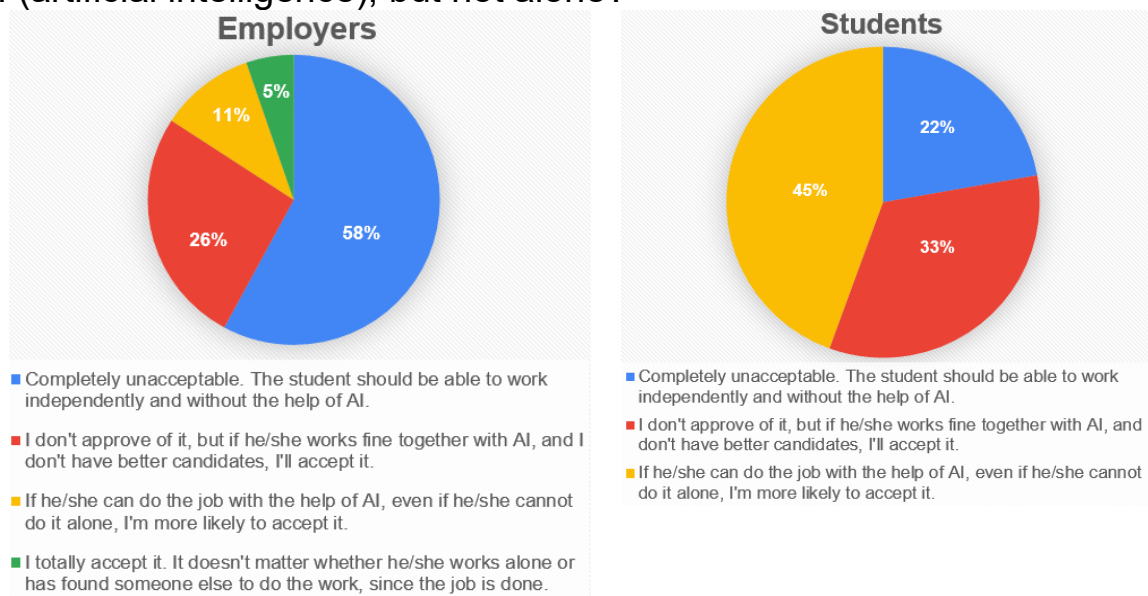

Fig. 3. According to 84% of employers it is not acceptable that students can write code with the help of AI, but not alone. However, 45% of students accept it.

There is a large discrepancy in the opinions of employers and students on this issue. *According to 84% (58%+26%) of employers, it is not acceptable that students can write code with the help of AI, but not alone. However, almost half of the students accept it.*

Another interesting question is Q13 in table 4. Obviously here students are even stricter than employers.

Table 4. Employers and students attitude towards using AI during an exam. Respondents were also able to choose "Good, C", "Very good, B" and "Excellent, A"

| Q13. How would evaluate a student who during a practical exam has to create an application, has the right to use all Internet resources, without AI, but he/she cannot do anything alone. However, if he/she is allowed to use artificial intelligence, then he/she gets an excellent solution from the AI, but does not understand it. | | |
|---|---|---|
| | **Employers** | **Students** |
| With a "Fail, F", because despite the ability to use everything on the Internet (without AI), he doesn't even know where to start | 47 % | 56 % |
| With a "Sufficient, D" or maximum "Good, C", because there is a working solution, however the student does not understand it | 53 % | 44 % |

The same question is given to the three most commonly used AI chatbots – ChatGPT [5], Claude [6] and DeepSeek [7]. Here is their answer:

Claude replies that in its opinion the student deserves Fail F, because the essence of both education and assessment is to measure students' knowledge and skills, not their ability to find someone/something to do the work for them.

ChatGPT states that the grade should be "Sufficient D", because we should not only evaluate knowledge, but also the ability to solve problems. If the student can solve problems with the help of AI, then he/she has at least some ability to find a solution.

According to DeepSeek the fair grade is "Sufficient D" since the problem is "solved", but the lack of understanding is a critical flaw in the educational context. The chatbot adds if the student can demonstrate how he/she used the AI, this shows some metacognitive skills and would even justify "Good C" grade.

**CONCLUSION**

After analyzing the results of the survey, we can conclude that:

1. For employers, *the algorithmic and logical thinking of graduates is extremely important*, while for almost half of the graduates themselves it is not the most important thing they need to learn at university.

2. Both employers (not all) and students believe that graduates should know at least one framework for the programming language being studied, but given the limited hours of workshops, if they had to choose, both groups of respondents overwhelmingly agree that *good knowledge of the programming language itself is more important* than knowing a framework or tools for automation and rapid development.

3. According to 2/3 of employers and half of students, the exam in subjects related to web programming should be *practical on a computer*.

4. Both employers and students believe that t*he learning process should employ and* rely *on tools and development environments with integrated artificial intelligence (AI)*, so that students get familiar with their features.

5. Almost half of students think *it acceptable for graduates to be able to write code with AI, but not alone*. While *84% of employers find this unacceptable*!

6. According to students, employers and artificial intelligence (ChatGPT, Claude and DeepSeek), if a student cannot cope on his own during an exam, but only with the help of artificial intelligence (AI), then the student should be assessed with a maximum grade of "Sufficient, D", because there is a working solution, but the student does not understand how it works.